\documentclass[sn-nature]{sn-jnl}% Style for submissions to Nature Portfolio journals
\usepackage{graphicx}%
\usepackage{multirow}%
\usepackage{amsmath,amssymb,amsfonts}%
\usepackage{amsthm}%
\usepackage{mathrsfs}%
\usepackage[title]{appendix}%
\usepackage{xcolor}%
\usepackage{textcomp}%
\usepackage{manyfoot}%
\usepackage{booktabs}%
\usepackage{algorithm}%
\usepackage{algorithmicx}%
\usepackage{algpseudocode}%
\usepackage{listings}%
\usepackage{inconsolata}
\usepackage{graphicx}
\usepackage{overpic}
\usepackage{multirow}
\usepackage{booktabs}
\usepackage{algorithm}
\usepackage{algpseudocode}
\usepackage[switch]{lineno}
\usepackage{tcolorbox}
\usepackage{wrapfig}

\usepackage{times}
\usepackage{amssymb}
\usepackage{placeins}
\usepackage{pifont}
\usepackage{float}
\usepackage[position=t,singlelinecheck=off,labelformat=empty,font=normalsize]{subfig} 
\theoremstyle{thmstyleone}%
\theoremstyle{thmstyletwo}%

\theoremstyle{thmstylethree}%

\begin{document}

\title[]{Leveraging LLM-based agents for social science research: insights from citation network simulations}

\author[1]{\fnm{Jiarui} \sur{Ji}}\email{jijiarui@ruc.edu.cn}
\author[1]{\fnm{Runlin} \sur{Lei}}
\author[3]{\fnm{Xuchen} \sur{Pan}}
\author*[1]{\fnm{Zhewei} \sur{Wei}}\email{zhewei@ruc.edu.cn}
\author[1]{\fnm{Hao} \sur{Sun}}
\author[1]{\fnm{Yankai} \sur{Lin}}
\author[1]{\fnm{Xu} \sur{Chen}}
\author[2]{\fnm{Yongzheng} \sur{Yang}}
\author[3]{\fnm{Yaliang} \sur{Li}}
\author[3]{\fnm{Bolin} \sur{Ding}}
\author[1]{\fnm{Ji-Rong} \sur{Wen}}

% \affil*[1]{\orgdiv{Gaoling School of Artificial Intelligence}, \orgname{Renmin University of China}, \orgaddress{\state{Beijing}, \country{China}}}
% \affil[2]{\orgdiv{School of Public Administration and Policy}, \orgname{Renmin University of China}, \orgaddress{\state{Beijing}, \country{China}}}
\affil*[1]{\orgdiv{Gaoling School of Artificial Intelligence}, \orgname{Renmin University of China}, \orgaddress{\state{Beijing}, \country{China}}}
\affil[2]{\orgdiv{School of Public Administration and Policy}, \orgname{Renmin University of China}, \orgaddress{\state{Beijing}, \country{China}}}

\affil[3]{\orgname{Alibaba Group},\country{China}}


\abstract{The emergence of Large Language Models (LLMs) demonstrates their potential to encapsulate the logic and patterns inherent in human behavior simulation by leveraging extensive web data pre-training. However, the boundaries of LLM capabilities in social simulation remain unclear. 
To further explore the social attributes of LLMs, we introduce the CiteAgent framework, designed to generate citation networks based on human-behavior simulation with LLM-based agents. CiteAgent successfully captures predominant phenomena in real-world citation networks, including power-law distribution, citational distortion, and shrinking diameter. 
Building on this realistic simulation, we establish two LLM-based research paradigms in social science: LLM-SE (LLM-based Survey Experiment) and LLM-LE (LLM-based Laboratory Experiment). 
These paradigms facilitate rigorous analyses of citation network phenomena, allowing us to validate and challenge existing theories. Additionally, we extend the research scope of traditional science of science studies through idealized social experiments, with the simulation experiment results providing valuable insights for real-world academic environments. Our work demonstrates the potential of LLMs for advancing science of science research in social science.}


\maketitle

\section{Main}

\begin{figure}[h]
  \centering
  \includegraphics[height=\linewidth, angle=270]{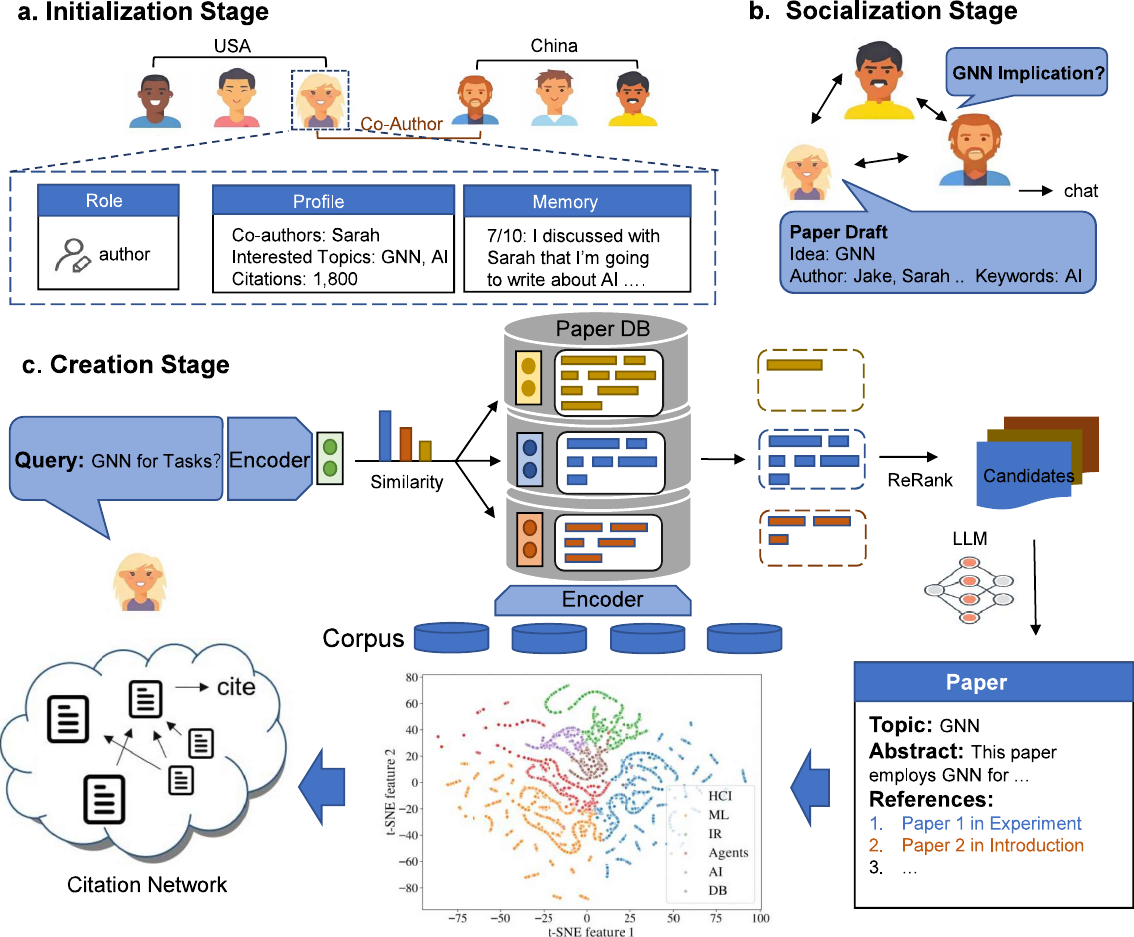}
  \caption{
    \textbf{An Illustration of One Simulation Step in the CiteAgent Framework.}
    %, mainly composed of three stages: Initialization, Socialization, and Creation.
  \textbf{a,} Initialization: LLM-based agents are built as distinct authors, each endowed with distinct attributes, denoted as $A$;
  \textbf{b}, Socialization: We designate an active author subset, denoted as $A_a$. Each active agent $a \in A_a$ engages in a group discussion with collaborators and collaboratively develops paper drafts.
  \textbf{c}, Creation: Each active agent utilizes a scholarly search engine to retrieve relevant papers and finalize the paper drafts with reference selection. 
  % \( n_p \) finished papers are subsequently added to \( P \).
  }
  \label{fig1}
\end{figure}

% Traditional studies of citation networks are conducted with random graph models, such as the Erdős–Rényi, Barabási–Albert, and small-world networks~\citep{klemm2002highly,centola2010spread}.
% Despite the simplicity and tractability of these models, they reduce human behavior to fixed statistical mechanisms (such as preferential attachment or local clustering), overlooking the heterogeneity, context dependence, and strategic adaptation observed in real-world settings~\citep{radicchi2011citation,koseoglu2016mapping}. 
% This limitation restricts the realism and generalizability of the resulting insights.
The science of science (SciSci) is an emerging subfield and interdisciplinary branch of the social sciences. Inheriting social science research methodologies, it systematically studies the processes, dynamics, and broader implications of scientific research itself~\citep{fortunato2018science}.
By investigating these mechanisms, SciSci offers critical insights into how science evolves as a collective enterprise, with potential implications for research efficiency, accelerating innovation and understanding collaboration patterns~\citep{scisci_cn}.
To achieve these goals, SciSci research employs diverse quantitative methods, with network science as the cornerstone for modeling scientific ecosystem evolution~\citep{newman2001structure}. 
Citation networks, in particular, provide a system-level perspective on the processes of the ideas, theories, and results spreading in science~\citep{radicchi2011citation}.

Traditional citation network studies are conducted with random graph models, such as the Erdős–Rényi, Barabási–Albert, and small-world networks~\citep{klemm2002highly,centola2010spread}.
Despite the simplicity and tractability of these models, they reduce human behavior to fixed statistical mechanisms (e.g., preferential attachment or local clustering), overlooking the heterogeneity and strategic adaptation observed in real-world settings~\citep{radicchi2011citation}. 
This restricts the realism and generalizability of the resulting insights.

Fortunately, Large Language Models (LLMs) like ChatGPT and GPT-4~\citep{anil2023palm,openai2024gpt4technicalreport} are emerging as promising tools. 
Through extensive web data pre-training, LLMs progressively show their potential to capture the logic and patterns necessary for simulating human behavior~\citep{gao2024large}. 
To address the limitations of conventional network models, we propose the \textbf{CiteAgent framework}, a simulation framework that leverages LLM-based agents to model human behaviors in citation networks. CiteAgent allows researchers to test, validate, or challenge established theories in network science with customizable experiment platforms.
% While network science has provided valuable structural insights, the behavioral assumptions underlying these models remain a critical limitation. 
% An increasing number of studies~\citep{de2023emergence, papachristou2024network, ji2024dynamic} come to integrate the role-playing capabilities of LLMs for social network generation. However, no research has yet employed simulation platforms to confirm or refute established theories in network science.

CiteAgent includes two entity types: the author set $A = {a_1, \dots, a_i, \dots a_n}$ and the paper set $P= {p_1, \dots, p_j, \dots p_m}$. 
We employ LLM-based agents to initiate authors and mimic human behaviors. We model interactions between these entities to trace the evolution of citation networks.
We initialize the author set $A$ and paper set $P$ from the citation network dataset $G_0$, which is enriched with seven paper attributes (e.g., \textit{paper content (abstract), paper timeliness, author name, author country, paper citation, paper topic, and author citation}) and five author attributes (e.g., \textit{citation, institution, expertise, topic, and nationality}).
Following the $G_0$ setup, we run iterative simulation steps in CiteAgent, as shown in Figure~\ref{fig1}. Each step integrates new papers and authors into $A$ and $P$ following three main stages: \textbf{Initialization, Socialization, and Creation.} In the initialization stage, we build LLM-based agents to instantiate authors. We update $A$ with new authors, and adjust for overloaded authors who have published excessively. Moving on to the socialization stage, authors collaborate, sharing insights and develop paper drafts. Finally, in the creation stage, authors finalize drafts by referencing relevant papers and updating $P$ with completed papers.
As simulation steps proceed, \(G_0\) progressively expands into a larger network. These steps represent the evolutionary process of the citation network. 
For validating the simulation authenticity of CiteAgent, we verify that the generated citation network conforms to several classic network science theories, including the power-law distribution~\citep{alstott2014powerlaw}, citational distortion~\citep{gomez2022leading} and shrinking diameter~\citep{leskovec2007graph} phenomenon.

Inspired by experimental social science methodologies, we propose two LLM-based experiment paradigms for studying human reference behaviors in SciSci research:

\textbf{(1) LLM-SE (LLM-based Survey Experiments)}:
Building on traditional survey experiments in which humans answer structured questionnaires~\citep{mullinix2015generalizability}, LLM-SE prompts questionnaires to LLM agents to analyze human reference behavior. The paradigm has three steps: 
(1) Variable Definition: define the independent and dependent variables. For example, how paper attributes (e.g., content, citation count) affect authors' citation inclination; 
(2) Questionnaire Design: construct prompts that embed variables to investigate LLM decision-making in citation contexts; 
(3) Quantitative Analysis: quantitatively analyze survey results to uncover LLM-based agent behavioral patterns.
\smallskip

\textbf{(2) LLM-LE (LLM-based Laboratory Experiments)}:
Building on traditional laboratory experiments that manipulate independent variables in controlled settings to infer causality~\citep{diamond1986laboratory}, LLM-LE designs different experimental conditions to isolate their effects on human reference behaviors. The paradigm has three steps: 
(1) Variable Definition: similar to LLM-SE, we need to define the independent and dependent variables for the experiment. For example, how different recommendation algorithms of scholarly search engines affect authors' citation behaviors;
(2) Experiment Design: construct baseline and treatment experiments that differ only in the presence or level of the independent variable, holding other factors constant, to isolate its causal effect;
(3) Causal Analysis: analyze the experiment results to derive causal insights into how specific experimental conditions shape citation patterns.

The CiteAgent framework provides a scalable and reproducible environment for studying human reference behaviors in network science. 
It supports controlled, large-scale simulations that complement traditional observational analyses by enabling systematic exploration of human behaviors under varying conditions. 
The proposed LLM-based experiment paradigms, LLM-SE and LLM-LE, offer a structured approach to simulate and analyze citation decisions, facilitating hypothesis-driven investigations for SciSci research.

\section{Results}

To demonstrate the application of CiteAgent in the science of science research, we employ three distinct datasets: CiteSeer, Cora~\citep{CiteGraph_Sen_08}, and LLM-Agent. 
The first two datasets are well-established in citation network analysis. To obtain text-rich data for each graph node, we crawl for abstract, author information, etc. The number of nodes in the CiteSeer dataset is 2,997, the Cora dataset is 2,542, and the LLM-Agent dataset is 165.
Furthermore, to study the evolution of the network from the beginning, we collect a new dataset consisting of all papers related to the emerging field of LLM-based agents, covering the period from 2021 to September 2023. 
The collected text-rich author attributes include five categories: \textit{citation, institution, expertise, topic, and nationality.} While for papers, the text-rich attributes include seven categories: \textit{paper content (abstract), paper timeliness, author name, author country, paper citation, paper topic, and author citation.}
We set the simulation start date to May 1, 2023, for the LLM-Agent dataset, and to January 1, 2004, for the CiteSeer and Cora datasets. Dataset collection procedures are described in Section 6.A of the supplementary material. These datasets serve as seed networks for generating the citation network. For CiteAgent construction, we select three top-ranked LLMs to explore the patterns of different LLMs in human behavior simulation: \texttt{GPT-3.5}, \texttt{GPT-4o-mini}~\citep{openai2024gpt4technicalreport} and \texttt{LLAMA-3-70B}~\citep{dubey2024LLAMA}.

\subsection{Power-law Distribution}

\begin{figure}[h]
  \subfloat{\includegraphics[width=0.95\linewidth]{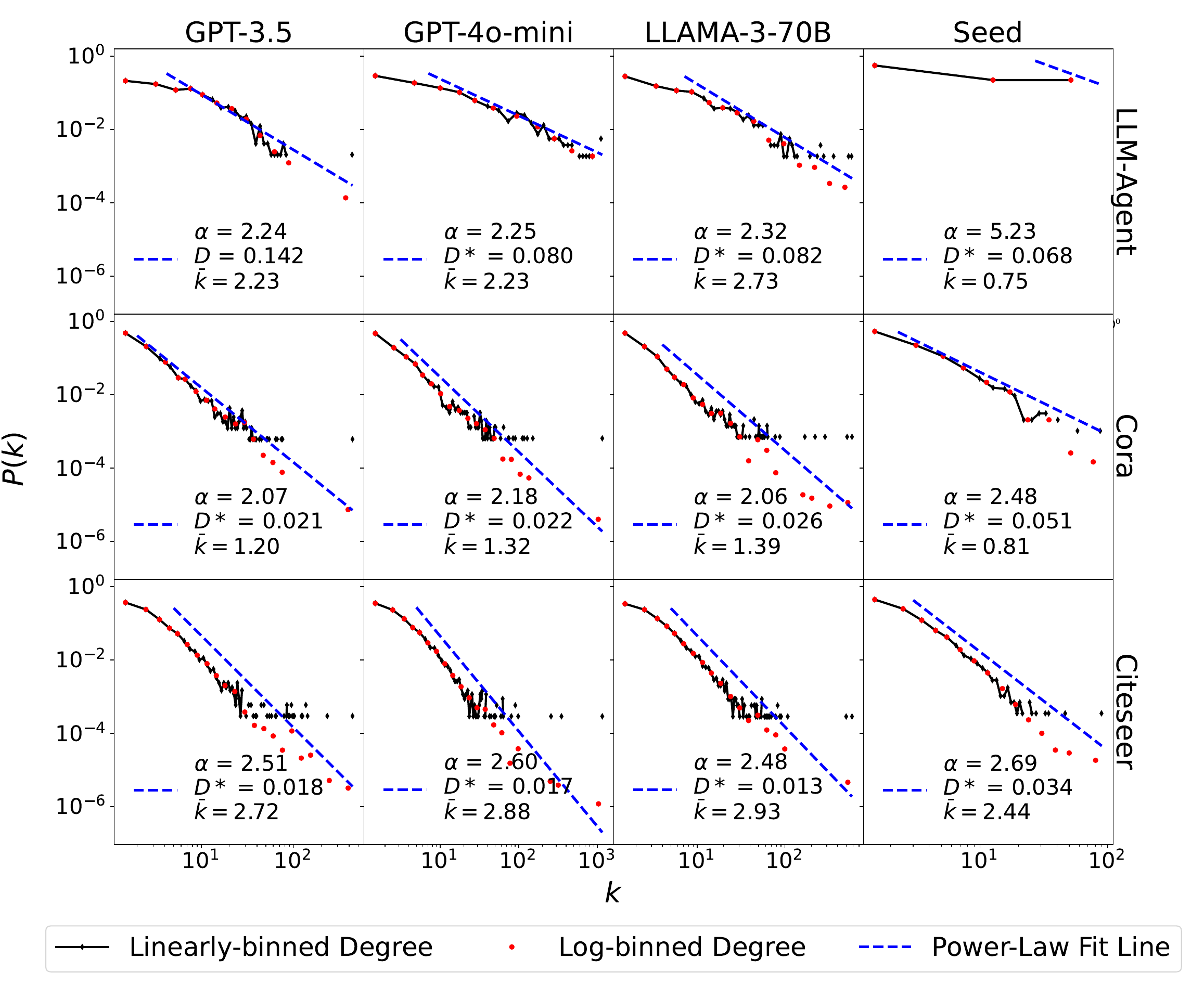}}
  \vspace{1em}
  \caption{\textbf{Power-Law Distribution in Citation Network Generated by CiteAgent}. CiteAgent expands the large dataset Cora and CiteSeer to 5000 nodes, and the LLM-Agent dataset to 1000 nodes. 
  We fit the network in-degree ($k$) distribution and with the power-law distribution model. $k$ is plotted in log-binned and linearly-binned formats against the probability density function $P(k)$ on a log-log scale. $\alpha$ indicates the power-law exponent, $D^*$ denotes a significant level at p-value $ < 0.01$, $\bar k$ indicates the average $k$.}
  \label{fig2}
\end{figure}

The degree distribution of citation networks often follows a \textit{power-law distribution}~\citep{barabasi1999emergence}, indicating a scale-free characteristic. Let $k$ represent the list of citation numbers received by an paper. The quantity $k$ obeys a discrete power-law model if it is drawn from a probability distribution:
$$
p(k) = \frac{k^{-\alpha}}{\zeta\left(\alpha, k_{\min }\right)},
$$
where \( \alpha\) is a shape parameter characterizing the distribution, known as the power-law exponent or scaling parameter and $\zeta\left(\alpha, k_{\min }\right)$ is the generalized or Hurwitz zeta function.
To assess the simulation authenticity of CiteAgent, we aim to verify whether the generated citation network obeys a power-law distribution. 
% And we use the real citation network datasets as the seed networks for a generation.

We employ the Kolmogorov-Smirnov statistic between the theoretical power-law model and the in-degree distribution, denoted as $D$, as an evaluation metric for fitness quantification, $D=\max_{k \geq k_{\text {min }}}|F(k)-H(k)|$.
Here $F(k)$ is the Cumulative Distribution Function (CDF) of the in-degree data with a value at least $k_{\text {min }}$, and $H(k)$ is the CDF for the power-law model that best fits the data in the region $k \geq k_{\text {min }}$. We follow the method outlined by~\cite{alstott2014powerlaw}, using \( D \) as the optimization objective to calculate cut-off in-degree \( k_{\text{min}} \). Additionally, we impose a constraint on the standard deviation of the $\alpha$, denoted as \( \sigma(\alpha)  < 0.1\) to ensure the stability of the power-law fitting. 
We adopt the \textit{Goodness-of-Fit} test for quantifying the plausibility of the power-law hypothesis~\citep{Clauset_2009}.
Specifically, if the resulting $D$ is less than the critical threshold corresponding to a p-value of 0.1, the power-law is a plausible hypothesis for the data, otherwise it is rejected.
For experiment setup, the recommended number of citations for each author is set to 3 in expanding the LLM-Agent dataset; For the Cora and CiteSeer datasets, the recommended number of citations falls within the range \([\bar{k}-\sigma(k), \bar{k}+\sigma(k)]\), where $k$ is the in-degree list of seed network, $\bar{k}$ is the average in-degree, and $\sigma(k)$ is the standard deviation of $k$.

As shown in Figure~\ref{fig2}, the expanded citation networks of the CiteSeer and Cora datasets can all closely fit a power-law distribution. 
We further examined the power-law fitness using randomized seed graphs of the CiteSeer and Cora datasets, as detailed in Section 5.B of the supplementary material. After graph expansion, we found that the in-degree distributions of networks generated from these randomized seeds still conform to a power-law model.
However, we find that in the LLM-Agent dataset, the expanded citation networks with \texttt{GPT-3.5} based authors cannot perfectly fit a power-law distribution, with $D = 0.142$ (p-value $> 0.01$), larger than expanded citation networks with \texttt{GPT-4o-mini} and \texttt{LLAMA-3-70B} based authors. This indicates that the latter two LLMs are more effective in simulating the human academic activities compared to \texttt{GPT-3.5}, and thus the expanded citation networks can better align with a power-law distribution.

\subsection{Reasons Behind Power-Law Distribution}

% \begin{figure}[htbp]
%   \centering
%   \subfloat[\textbf{a}]   
%     {
%         \centering
%         \includegraphics[width=.48\linewidth]{figures/Fig3a.pdf}
%         \label{fig3}-a
%     }
%   \subfloat[\textbf{b}]  
%     {
%         \centering
%         \includegraphics[width=.5\linewidth]{figures/Fig3b.pdf}
%         \label{fig3}b
%     }
%     \\
%     \subfloat[\textbf{c}] 
%     {
%         \centering
%         \includegraphics[width=\linewidth]{figures/Fig3c.pdf}
%         \label{fig:reason_llmagent_ks}c 
%     }
%     \caption{\textbf{The LLM-LE and LLM-SE Analysis for different LLMs in Forming Power-Law Distributions.}
%     \textbf{a,} LLM-LE: $D$ for generated citation networks in all experimental conditions.
%     \textbf{b,} LLM-LE: maximum in-degree for citation networks in all experimental conditions.
%     \textbf{c,} LLM-SE: the predominant paper influencing author reference selection behavior.
%     }

% \end{figure}

\begin{figure}[htbp]
\centering
  \includegraphics[width=\linewidth]{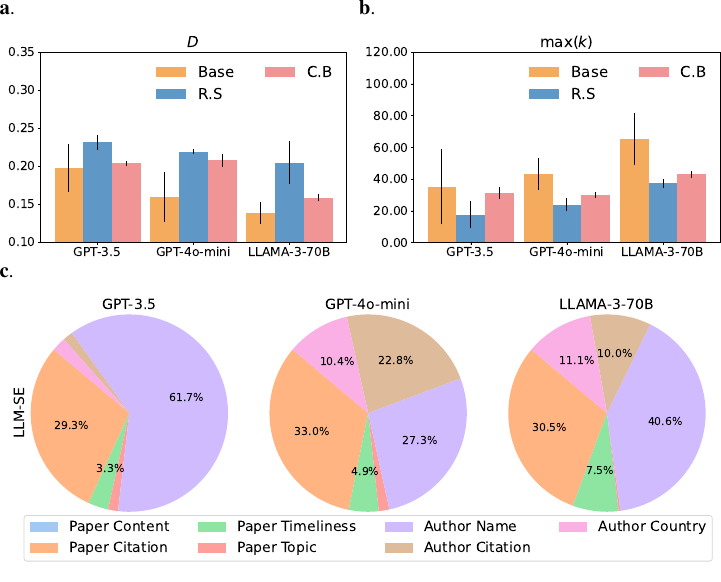}
  
  \caption{\textbf{The LLM-LE and LLM-SE Analysis for different LLMs in Forming Power-Law Distributions.}
    \textbf{a,} LLM-LE: $D$ for generated citation networks in all experimental conditions.
    \textbf{b,} LLM-LE: maximum in-degree for citation networks in all experimental conditions.
    \textbf{c,} LLM-SE: the predominant paper influencing author reference selection behavior.
    }
  \label{fig3}
\end{figure}

The most well-known explanation for the power-law distribution is preferential attachment~\citep{borner2007network}. However, the exact mechanisms underlying preferential attachment remain uncertain. 
According to the Barabási-Albert model~\citep{barabasi1999emergence}, the probability of the paper being cited is proportional to the current citation count. 
Another theory, based on heterogeneous growth rates~\citep{krapivsky2000connectivity}, suggests that the likelihood of a paper being cited depends on its age and timeliness. 
Additionally, some authors have claimed that citations can be influenced by the nationality and institution of the authors~\citep{gomez2022leading}.

% \textbf{So, what impacts the reference selection of LLMs and cultivates power-law distribution?}
\textbf{Factors Influencing Reference Selection of LLM-based Authors.}
We explore the reference selection pattern of LLM-based authors with LLM-SE and LLM-LE paradigms.

At first, we adopt LLM-SE to identify the predominant paper attribute for LLMs in reference selection. In this LLM-SE experiment, we define independent variables as paper attributes (e.g., \textit{paper content, paper timeliness, paper citation, paper topic, author name, author country, and author citation}), the dependent variable as the power-law distribution fitness of the generated citation network.
Then we design a structured prompt that lists seven options (paper content, timeliness, citation count, topic, author name, country, and author citation), requiring agents to identify the single most influential factor in their reference selection.  
Lastly, we conduct quantitative analysis of the LLM-SE result. As illustrated in Figure~\ref{fig3}c, compared to \texttt{GPT-3.5}, the bias observed in \texttt{LLAMA-3-70B} and \texttt{GPT-4o-mini} primarily stems from citation-related information (e.g., paper citation, author citation, and paper timeliness).
Consequently, LLM-based agents using these models are more likely to engage in preferential attachment to highly cited papers, leading to citation networks with a more pronounced power-law distribution.

We further apply LLM-LE to interpret the factors contributing to the power-law distribution observed in the generated citation networks. Specifically, in this LLM-LE experiment, we employ three experimental conditions to isolate causal factors. First, we designate two independent variables, including the recommendation algorithm and citation information visibility, and the dependent variable is defined as the power-law distribution fitness of the generated citation networks.
Next, we formulate three experimental conditions to isolate causal factors for power-law fitness of generated networks:

\textbf{(1) Base:} Control condition. Papers are retrieved based on maximum embedding similarity to the query $Q_i$. The top 20 most cited papers among the results are selected to form the candidate set $P_i^{c}$. 
All paper content remains visible.

\textbf{(2) Random Search (R.S):} Experimental condition with random paper recommendation. We retrieve 100 relevant papers using $Q_i$, then randomly select 20 papers to constitute $P_i^{c}$. 
All paper content remains visible.

\textbf{(3) Citation Blind (C.B):} Experimental condition with obscured citation information. While the paper recommendation algorithm is kept constant, all citation-related content (e.g., paper citation, author citation) is masked.

Lastly, we conduct a causal analysis of the LLM-LE results. Due to the high computational cost of simulations, we repeat the experiments on all LLM-based agents three times. In network science, a hub is a node with a significantly higher degree than others, often emerging due to the preferential attachment mechanism~\citep{krapivsky2000connectivity}. In citation networks, highly cited papers that receive numerous references naturally evolve into hubs. As shown in Figure~\ref{fig3}b, large hubs ($max(k)$) are more likely to emerge in the Base experiment, indicating that the preferential attachment mechanism is more pronounced.
We set \( k_{\text {min }} = \max(k) \times 0.05 \) and compute the \( D \) for all citation network in-degree distributions.
As shown in Figure~\ref{fig3}a, we observe the following:

First, in simulation experiments with different LLM-based authors, randomly recommending papers consistently decreases the power-law fitness of in-degree distributions and reduces the in-degree of hubs, as evidenced by an increase in $D$ and a drop in $max(k)$. This effect is due to the search engine's role in defining the candidate papers visible to authors. Specifically, by reducing the probability of recommending highly cited papers, the likelihood of authors' preferential referencing decreases, thereby lowering the fitness of the network's in-degree distribution to the power-law distribution.

Second, random paper recommendations negatively affect the power-law fitness of the in-degree distribution more than anonymizing citation-related information, suggesting that the recommendation algorithm reducing the likelihood of recommending highly cited papers is a stronger driver of preferential attachment than citation bias.

Lastly, anonymizing citation-related information lowers power-law fitness in simulations with \texttt{LLAMA-3-70B} and \texttt{GPT-4o-mini} based authors, indicating their inclination to preferentially reference highly cited papers.
Conversely, in the simulations of \texttt{GPT-3.5} based authors, we find no substantial decline in power-law fitness and the in-degree of hub. 
This finding based on LLM-LE experiments is consistent with the insights from LLM-SE experiments, which indicate that \texttt{GPT-3.5} based authors exhibit a lack of preferential referencing to highly cited papers. 
This suggests that as LLM capabilities improve, more advanced models (such as \texttt{LLAMA-3-70B} and \texttt{GPT-4o-mini}) more effectively simulate human values and behaviors, such as preferential referencing, compared to less capable models like \texttt{GPT-3.5}.

\subsection{Citational Distortion}

% \begin{figure}[h]
%   \centering
%   \subfloat[\textbf{a}] 
%     {
%         % \centering
%         \includegraphics[width=.48\linewidth]{figures/Fig4a.pdf}
%         \label{fig4}a
%     }
%   \hspace{30pt}
%   \subfloat[\textbf{b}]  
%     {
%         \centering
%         \includegraphics[width=.35\linewidth]{figures/Fig4b.pdf}
%         \label{fig4}b
%     }
%  \\
%   \subfloat[\textbf{c}]  
%     {
%         \centering
%         \includegraphics[width=.98\linewidth]{figures/Fig4c.pdf}
%         \label{fig4}c
%     }
%     \\
    
%   \caption{\textbf{The LLM-SE and LLM-LE Analysis for the Citational Distortion Phenomenon.} 
%   \textbf{a,} LLM-SE: the reference selection proportion driven by country-related information, grouped by papers from different countries. 
%   \textbf{b}, The citational distortion result figure from~\citep{gomez2022leading}. 
%   \textbf{c}, LLM-LE: the $\beta$ evolution trend in different experimental conditions, which shows that preferential attachment causes $\beta$ coefficient exceeding.}
% \end{figure}

\begin{figure}[htbp]
\centering
  \includegraphics[width=\linewidth]{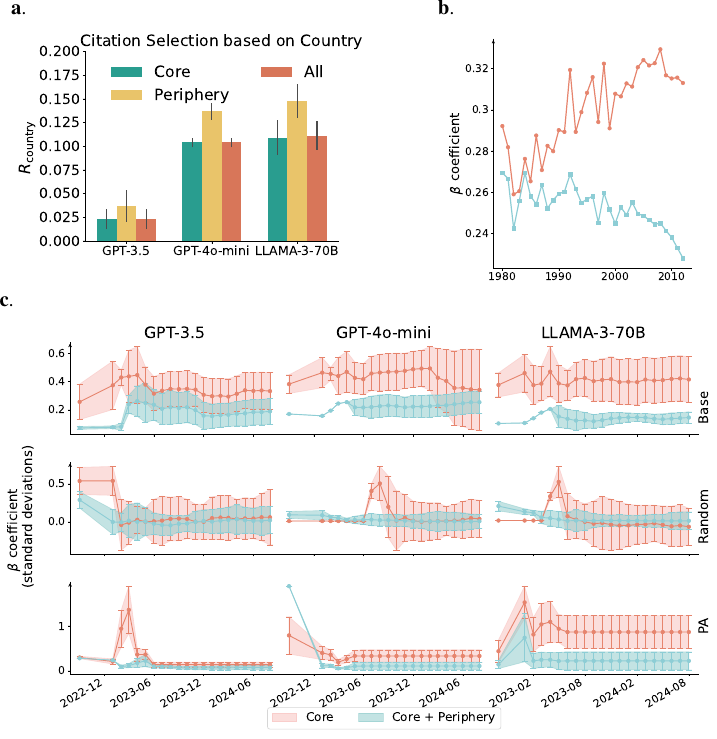}
  \caption{\textbf{The LLM-SE and LLM-LE Analysis for the Citational Distortion Phenomenon.} 
  \textbf{a,} LLM-SE: the reference selection proportion driven by country-related information, grouped by papers from different countries. 
  \textbf{b}, The citational distortion result figure from~\citep{gomez2022leading}. 
  \textbf{c}, LLM-LE: the $\beta$ evolution trend in different experimental conditions, which shows that preferential attachment causes $\beta$ coefficient exceeding.}
  \label{fig4}
\end{figure}

In addition to analyzing the citation network structure, we also examine the citation pattern of LLM-based authors from the network textual attributes. Specifically, we focused on the citational distortion phenomenon proposed in~\citep{gomez2022leading}, which suggests that paper citation can be influenced by the nationality of the authors, with the core countries \(C_{core}\) increasingly receiving more citations than peripheral countries \(C_{peripheral}\). 

They investigate this phenomenon with citational lensing framework, which uses a network regression model called the semi-partialing quadratic assignment procedure (QAP)~\citep{dekker2007sensitivity} to capture the extent to which citations are associated with paper content similarity with the regress coefficient $\beta$ coefficient. They claim that the $\beta$ coefficient exceeding phenomenon, where $\beta$ is consistently higher for core countries than for all countries, indicates the existence of the citational distortion.

To replicate the analysis process in~\citep{gomez2022leading}, we first classify countries based on paper counts across countries using the CiteSeer and Cora datasets (with country determined by author profiles). The top 50\% are designated as core countries, and the remainder as peripheral countries. Next, we initialize \( G_0 \) with the LLM-Agent dataset, expanding it to 500 nodes. Finally, we calculate the \( \beta \) coefficient with citation lensing framework.

\textbf{The cumulative advantage drives citational distortion.}
To investigate the $\beta$ coefficient exceeding phenomenon in citational distortion, we conduct an LLM-SE experiment to analyze country-level citation biases in CiteAgent. In this LLM-SE experiment, we define independent variables as \textit{author country} (core vs. peripheral countries), with the dependent variable as the proportion of citations most influenced by \textit{author country}, denoted as $\operatorname{R_\text{country}}$;  
Then, we prompt LLM-based agents from core and peripheral countries with a structured questionnaire. We ask them to identify the most influential factor in their reference selection. Options include \textit{paper content, paper timeliness, paper citation, paper topic, author name, author country and author citation.} Finally, we get quantitative analysis of the LLM-SE result. As illustrated in Figure~\ref{fig4}a, we are surprised to find that \(\operatorname{R_\text{country}}\) for core countries is comparable to, or even slightly lower than, that for peripheral countries. This finding suggests the authors' intentional bias toward papers from core countries is uncertain, and thus the \(\beta\) coefficient may not serve as a distinctive indicator of intentional bias.

To further probe the drivers of the $\beta$ exceeding phenomenon, we design three experiments using LLM-LE. First, we designate one independent variables: the structure of the country-wise citation network, and the dependent variable is defined as the $\beta$ coefficient exceeding phenomenon. Next, we formulate three experimental conditions, varying the structure of the country-wise citation network to isolate its effect on \(\beta\) coefficient:

\textbf{(1) Base:} Control condition. The country-wise citation network and paper content network are generated by CiteAgent to compute the $\beta$ coefficient under empirically observed conditions.

\textbf{(2) Random:} Experimental condition with randomized country assignment. A country-wise citation network is constructed by uniformly sampling country labels for each paper, breaking the assumption that core countries produce more papers.

\textbf{(3) Preferential Attachment (PA):} Experimental condition based on preferential attachment. The country-wise citation network is generated such that a country’s probability of receiving a new citation is proportional to its cumulative paper count, amplifying accumulative advantages over time.

Lastly, we analyze the LLM-LE results. As shown in Figure~\ref{fig4}c, CiteAgent successfully simulates the citational distortion phenomenon in the Base experimental condition, with a higher $\beta$ coefficient for core countries than all countries (core plus peripheral). 
This is consistent with the trend presented in~\citep{gomez2022leading}, as shown in Figure~\ref{fig4}b.
However, the $\beta$ coefficient for core countries still exceeds that for all countries in the PA experimental condition, but vanishes when country labels are randomly reassigned in the Random environment.
The preferential attachment, which is the fundamental reason for power-law citation distribution in the citation network: newly added papers tend to cite highly cited hub papers. This amplifies existing disparities through \emph{cumulative advantage}~\citep{perc2014matthew}: countries with more papers are more likely to receive new citations, leading to a Matthew effect dynamic.
The result in the PA experimental environment indicates that preferential attachment alone constitutes a sufficient condition for the $\beta$ coefficient exceeding phenomenon claimed in~\citep{gomez2022leading}. 
Thus, an elevated $\beta$ for core countries can arise from network-growth dynamics rather than deliberate citing for papers from core countries over peripheral countries given similar paper content. Consequently, $\beta$ coefficient alone cannot reliably indicate country-wise citational distortion; at best, it reflects the influence of preferential attachment and the resulting structural inequality present in the citation network evolution.

\subsection{Reasons Behind Citational Distortion}

% \begin{figure}[htbp]
%   \centering
%   \begin{minipage}[t]{0.46\textwidth}
%     \subfloat[\textbf{a}]%Self Citation Rate by Country.]  
%     {
%         \includegraphics[width=\linewidth]{figures/Fig5a.pdf}
%         \label{fig5}a
%     }
%   \end{minipage}%
%   \quad
%   \begin{minipage}[t]{0.46\textwidth}
%     \subfloat[\textbf{b}]%Counterfactual Experiments.]   
%     {
%         \centering
%         \includegraphics[width=.93\linewidth]{figures/Fig5b.pdf}
%         \label{fig5}b
%     }
%     \\
%     \subfloat[\textbf{c}]%RPS for Citation Networks.]  
%     {
%         \centering
%         \includegraphics[width=.95\linewidth]{figures/Fig5c.pdf}
%         \label{fig5}c
%     }
%   \end{minipage}%
%   \\
%     \subfloat[\textbf{d}]
%     {
%         \centering
%         \includegraphics[width=.9\linewidth]{figures/Fig5d.pdf}
%         \label{fig5}d
%         \vspace{.1\linewidth}
%     }
    
%   \caption{\textbf{Examination and Analysis of Citational Distortion.} 
%   \textbf{a,} A comparison of self-citation rate (SCR) by country using the Scopus dataset, alongside the citation networks generated in the base and equal author experimental conditions.
%   \textbf{b}, A comparison of the Gini coefficient in public and anonymous experimental conditions. 
%   \textbf{c}, Comparison of RPS between core countries and peripheral countries in the public experimental condition.
%   \textbf{d}, RPS evolution process in the public experimental condition.
%   }
% \end{figure}

\begin{figure}[htbp]
\centering
  \includegraphics[width=\linewidth]{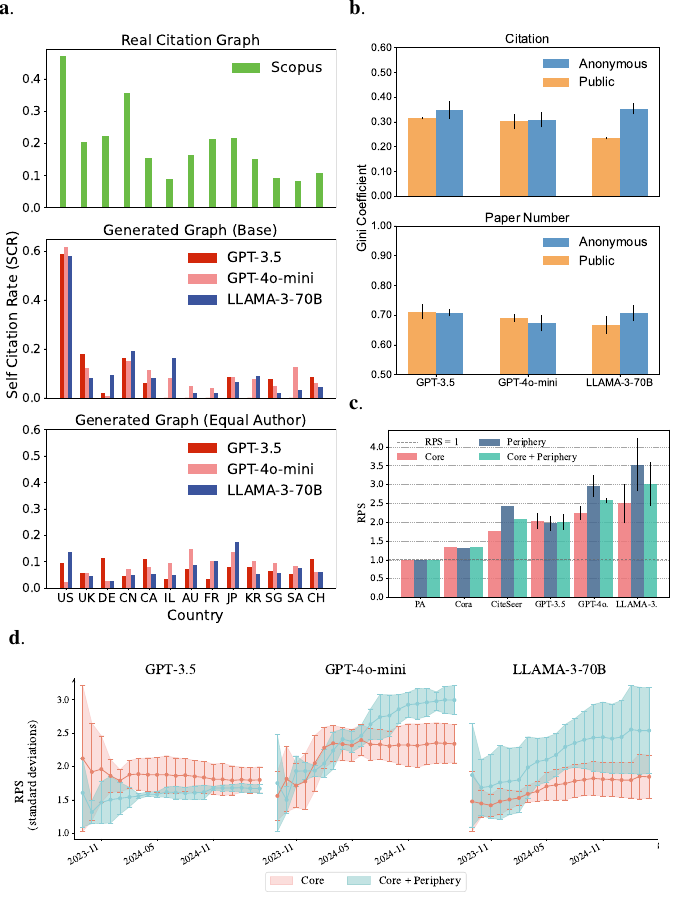}
  \caption{\textbf{Examination and Analysis of Citational Distortion.} 
  \textbf{a,} A comparison of self-citation rate (SCR) by country using the Scopus dataset, alongside the citation networks generated in the base and equal author experimental conditions.
  \textbf{b}, A comparison of the Gini coefficient in public and anonymous experimental conditions. 
  \textbf{c}, Comparison of RPS between core countries and peripheral countries in the public experimental condition.
  \textbf{d}, RPS evolution process in the public experimental condition.
  }
  \label{fig5}
\end{figure}

% \textbf{Assessing Whether Core Countries Benefit from Superior Referencing.}

Since the $\beta$ coefficient exceeding is insufficient to indicate that papers from core countries are prioritized in citations, the existence of citational distortion remains uncertain. 
The high self-citation rate (SCR) in core countries has often been interpreted as evidence of national preference~\citep{BARDEESI2021100333}. To examine country-specific citation patterns among authors, we conduct simulation experiments within CiteAgent. We begin by collecting author country information from the CiteSeer and Cora datasets, categorizing countries as core or peripheral based on paper counts.

We examine authors' referencing preferences for domestic versus international papers. Prior studies have found that SCR among countries show highly skewed distributions, with core countries exhibiting higher SCR in the Scopus dataset~\citep{BARDEESI2021100333}.
As illustrated in Figure~\ref{fig5}a, 
authors display a pronounced \textit{relative own-language preference}~\citep{yitzhaki1998language}. This preference is marked by a highly skewed SCR distribution both in real and citation networks generated in the Base environment. Core countries such as the United States, the United Kingdom, and China exhibit high SCR.

To investigate the reasons behind the skewed SCR distribution, we adopt LLM-LE to conduct causal analyses of two hypothesized mechanisms:
(1) preferential referencing of authors due to country information; 
(2) structural inequality arising from unequal author distribution across countries.

First, we examine whether author show preferential referencing due to country information. We follow the established indicator of Matthew effect. We validate the existence of citational distortion by applying the Gini coefficient (co-Gini)~\citep{fu2020fairness} to country-specific citation and paper numbers. We define the independent variables as country information, while the dependent variable is co-Gini.
Next, we design two experimental conditions, manipulating the visibility of author and country information:

\textbf{(1) Anonymous:} Author and country information are fully anonymized and hidden from the LLM author.

\textbf{(2) Public:} Author and country information are fully disclosed and accessible during the citation decision process.

Lastly, we conduct causal analysis of the LLM-LE results. As shown in Figure~\ref{fig5}b, increasing anonymity does not necessarily reduce the co-Gini coefficient for country-wise citations. Simulations with \texttt{GPT-3.5} and \texttt{LLAMA-3-70B} based authors show an increase in co-Gini values. In terms of country-wise paper numbers, the co-Gini remains relatively stable between anonymous and public environments. 

These findings suggest that anonymity does not promote a more equitable distribution of citations and paper numbers across countries. In other words, core countries are not prioritized in referencing given similar paper content, with and without country information. Given that the co-Gini results show no preferential referencing of core countries, we examine the second potential driver of the skewed SCR distribution. In this LLM-LE experiment, we define independent variables as the country-wise author number, the independent variable as country-specific SCR. We design two experimental conditions to compare SCR:

\textbf{(1) Base:} The nationality of the author is determined by LLM, where core countries are more likely to be chosen.

\textbf{(2) Equal Author:} The nationality of the author is chosen uniformly at random.

Lastly, we conduct causal analysis of the LLM-LE results. As shown in the last row of Figure~\ref{fig5}a, the SCR follows a uniform distribution when the number of authors is equal across countries. This indicates that the unequal distribution of authors by country, where core countries have more authors, contributes to the elevated SCR in core countries over peripheral countries.

% \textbf{Indicators of reference selection preference for country-specific papers.}
\textbf{Towards a more reliable indicator of the citational distortion.}
In core countries, the skewed SCR largely stems from an uneven distribution of authors. This imbalance leads both the  $\beta$ and the SCR indicator to conflate structural growth with intentional citational bias. As a result, neither metric reliably indicates intentional citational distortion.

To address this issue, we propose a new metric that normalizes for the cumulative advantage of paper volume. Inspired by the Institutional Preference Score (IPS), which was introduced to capture the citation preferences of various institutions for AI papers in the industry~\citep{frank2019evolution}. 
Likewise, we propose the Referencing Preference Score (RPS) to examine the referencing preferences of agent set \(A\) for papers from country set \(C\):

\begin{equation*}
  \begin{aligned}
    \text{RPS }_{\text {year}}(C, A) &= \frac{1}{|A|} \sum_{a}^{A} \frac{\text {(citation share of } C \text{ in citations of } a)}{\text{(paper share of } C)} \text{ from the beginning to year.}\\
    \text{RPS}(C) &= \sum_{year}^{}\frac{\text{RPS }_{\text {year }}(C, A)}{|\text{year} - \text{beginning}|}
    \label{eq:RPS}
  \end{aligned}  
\end{equation*}

Here, $\text{RPS}(C) > 1$ demonstrates a preference for papers from \(C\) than would be expected under random referencing behaviors. If reference selection is determined by preferential attachment based on paper counts, we would expect the referencing probability to mirror the distribution of these counts, resulting in an RPS of 1 for all countries (if the ref. share of \(C\) in papers is zero, we default to $\text{RPS}(C) = 1$. We repeat all simulation experiments three times to calculate the RPS standard deviation error.
As shown in Figure~\ref{fig5}c, for peripheral countries, we find that $\text{RPS}(C_\text{core})$ is comparable to, or even slightly lower than, $\text{RPS}(C_\text{peripheral})$. Moreover, as illustrated in Figure~\ref{fig5}d, in the simulations of \texttt{GPT-4o-mini} and \texttt{LLAMA-3-70B} based authors, $\text{RPS}(C_\text{peripheral+core})$ gradually exceeds $\text{RPS}(C_\text{core})$ over the years. In contrast, in the simulations of \texttt{GPT-3.5} based authors, \(\text{RPS}(C_{\text{peripheral+core}})\) consistently remains slightly lower than \(\text{RPS}(C_{\text{core}})\).
Thus, in the CiteAgent simulations, authors exhibit no significant preference for papers from peripheral countries over those from core countries; in fact, they show a slight inclination toward papers from peripheral countries. This observation is consistent with the citation motivation statistics presented in Figure~\ref{fig4}a and the results of anonymous-public counterfactual experiments, further confirming the absence of preference for papers from core countries over those from peripheral countries.
Therefore, the observed inequality in international citations primarily stems from the uneven distribution of researchers across countries, particularly the concentration of authors in core countries, rather than from intentional biases based on the authors' nationality during the reference selection.

\subsection{Idealized Social Experiment}

% \textbf{Exploring the Potential of CiteAgent to Expand Traditional SciSci Research.}
Based on the preferential referencing behavior of LLM-based authors, CiteAgent is capable of realistically simulating the evolution of citation networks. It also enables counterfactual experiments to analyze the underlying mechanisms of referencing behaviors. This motivates us to expand traditional SciSci research by designing idealized academic environments within CiteAgent.

% \begin{figure}[htbp]
%   \centering
%   \subfloat[\textbf{a}] %Traditonal CCA Analysis]   
%     {
%         \centering
%         \includegraphics[width=\linewidth]{figures/Fig6a.pdf}
%         \label{fig6}a
%     }
%     \\
%   \subfloat[\textbf{b}] %Traditonal CCA Analysis]  
%     {
%         \centering
%         \includegraphics[width=.6\linewidth]{figures/Fig6b.pdf}
%         \label{fig6}b
%     }
%   \subfloat[\textbf{c}] %LLM-SE-based CCA Analysis]  
%   {
%         \includegraphics[width=.4\linewidth]{figures/Fig6c.pdf}
%         \label{fig6}c
%     }
%     \\
%     \subfloat[\textbf{d}] %LLM-LE-based CCA Analysis] 
%   {
%     \includegraphics[width=.8\linewidth]{figures/Fig6d.pdf}
%         \label{fig6}d
%         % \hspace{.2\linewidth}
%     }
%   \caption{\textbf{Idealized Social Experiment.} 
%   \textbf{a,} A network evolution experiment demonstrating real-world network properties of densification and shrinking diameter, using \texttt{GPT-3.5} based authors for simulation.
%   \textbf{b,} Frequency analysis of cited papers based on referencing motivations and placements, compared against real data from empirical studies.
%   \textbf{c,} Importance score of cited papers categorized by different referencing motivations and placements, using \texttt{GPT-3.5} based authors for simulation.
%   \textbf{d,} The paper counts evolution trends in single-author and multi-author experimental conditions. }
% \end{figure}

\begin{figure}[htbp]
\centering
  \includegraphics[width=\linewidth]{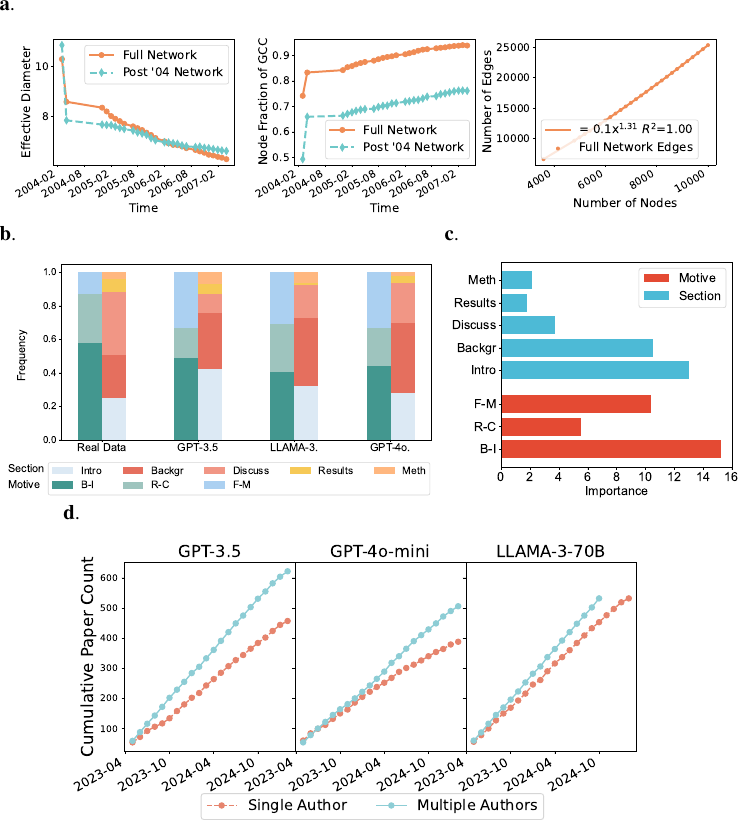}
  \caption{\textbf{Idealized Social Experiment.} 
  \textbf{a,} A network evolution experiment demonstrating real-world network properties of densification and shrinking diameter, using \texttt{GPT-3.5} based authors for simulation.
  \textbf{b,} Frequency analysis of cited papers based on referencing motivations and placements, compared against real data from empirical studies.
  \textbf{c,} Importance score of cited papers categorized by different referencing motivations and placements, using \texttt{GPT-3.5} based authors for simulation.
  \textbf{d,} The paper counts evolution trends in single-author and multi-author experimental conditions. }
  \label{fig6}
\end{figure}

\textbf{Citation network evolution experiment.}
In the past empirical study of citation network evolution,~\cite{leskovec2007graph} identify two surprising properties: densification, where the ratio of edges to nodes increases over time, and shrinking diameter, where the network's diameter decreases to a constant value over time~\citep{lattanzi2009affiliation}. To test whether CiteAgent can replicate these phenomena, we initialize agents with \texttt{GPT-3.5} and \( G_0 \) with the CiteSeer dataset. To mitigate any interference from \( G_0 \), we investigate the evolution patterns of the Full Network and Post '04 Network (which considers 2004 as the starting year for simulation). As shown in Figure~\ref{fig6}a, we observe the densification power law in the citation network's evolution, expressed as \( n(t) \propto e(t)^{\gamma}, \gamma = 1.31\), where \( e(t) \) and \( n(t) \) represent the edge and node counts in the graph at time \( t \), respectively. In addition, we confirm the shrinking diameter property. Using the \( q \)-effective diameter metric defined in~\citep{leskovec2007graph} with \( q = 0.9 \), we observe a consistent reduction in effective diameter over time across both the Full Network and the Post-'04 Network. We also measure the fraction of nodes within the largest connected component (GCC) over time, finding that it increases as the network evolves. The formation of a dominant GCC, combined with the densification power law, indicates that as the network matures, it becomes more densely interconnected. This densification contributes to the increasing interconnectedness in citation networks, corroborating the emergence of shrinking diameter phenomena~\citep{leskovec2007graph,lattanzi2009affiliation}.

\textbf{CCA analysis experiment.}
Traditional Citation Content Analysis (CCA) predominantly relies on extensive data collection~\citep{frank2019evolution} and hypothesis testing~\citep{zhang2012citationcontentanalysiscca}, focusing on statistical characteristics of referencing motivation and section~\citep{THELWALL2019658,cohan-etal-2019-structural}. In this study, we employ CiteAgent to replicate and extend this analytical framework through two LLM-SE experiments.
The first LLM-SE experiment investigates section placement in citation behavior. We define the independent variable as the set of possible sections: \textit{introduction, background, results, discussion}, and the dependent variable as the frequency and its importance. Following the schema of~\citep{THELWALL2019658}, we design a structured questionnaire that prompts LLM agents to select the most appropriate section for each citation and to evaluate the relative importance of each section in academic writing.
The second LLM-SE experiment examines citation motivations. We define the independent variable as the SciCite taxonomy categories~\citep{cohan-etal-2019-structural}: \textit{providing background information (B-I), describing methods (F-M), comparing results (R-C)}, and the dependent variable as the selected motivation and its importance. We design a structured prompt that asks LLM agents to identify the primary motivation for each citation and to relative importance of each motivation category.

Lastly, we conduct quantitative analysis of the LLM-SE result. As shown in Figure~\ref{fig6}b, different LLMs exhibit consistent patterns in citation motivations and section usage. The dominant motivation is providing background information (B-I), and citations are most frequently placed in the discussion, background, and introduction sections. Figure~\ref{fig6}c further reveals that, the most important referencing section is Background and Introduction, while the most important referencing motivation is providing background information.

% 待添加format

\textbf{Co-authorship ablation experiment.}
Past research has demonstrated the substantial influence of co-authorship on citation dynamics, with factors such as closeness centrality~\citep{biscaro2014co} and homophily~\citep{hancean2016homophily} within co-author networks shaping referencing behaviors. Nonetheless, conventional studies face limitations in experimentally manipulating co-authorship variables at scale. 
To address this limitation, we conduct contrastive LLM-LE experiments.
First, we define the independent variable as authorship configuration, operationalized as either single-author or multi-author collaboration in paper generation; while the dependent variable is publication volume, measured through the cumulative paper count. Next, we formulate two experimental conditions where we vary authorship configuration to isolate its effect on paper creation:

\textbf{(1) Single Author:} Each paper is written by one LLM-based agent.

\textbf{(2) Multi-Author:} Each paper is collaboratively written by multiple LLM-based agents, with the number of authors per paper ranging up to five.

Lastly, we conduct causal analysis of the LLM-LE results. Our findings highlight that single authorship diminishes the creativity of authors. As shown in Figure~\ref{fig6}d, there is a marked decrease in the volume of published papers in single-author simulations with \texttt{GPT-3.5}, \texttt{GPT-4o-mini}, and \texttt{LLAMA-3-70B} based authors. This decline stems from the repetitive content of papers in single-author simulations, which leads to a lack of novelty in the papers and ultimately reduces the number of papers. 

% 待添加format

\section{Discussion}

First, CiteAgent explores the capability boundaries of untrained LLMs in capturing human behavior patterns. Recently, there have been several works integrating LLMs into the field of human behavior simulation~\citep{shao2023character,ji2024srap}, typically rely on the Turing test to assess the authenticity of their simulations. However, a significant concern in this area is the reproducibility and generalizability of the evaluation metrics used. In contrast, CiteAgent validate simulation authenticity and reliability by aligning simulation results with established theories from social science like the power-law distribution~\citep{alstott2014powerlaw}, the citational distortion~\citep{gomez2022leading}, shrinking diameter~\citep{leskovec2007graph} etc.

Additionally, through our proposed research paradigms of LLM-LE and LLM-SE, CiteAgent is particularly meaningful in validating and challenging existing theories. Preferential attachment is one of the common explanations for power-law distributions. 
Our counterfactual experiments confirm that both the recommendation algorithm of scholarly search engines and the inclination of LLM-based authors to cite hub papers mutually reinforce preferential attachment, thereby contributing to the establishment of a stable power-law degree distribution in the evolution of citation networks.

Furthermore, regarding the citational distortion phenomenon observed in~\citep{gomez2022leading}, where papers from core countries consistently receive more citations for similar content, \(\beta\) is used as a metric to assess these citation disparities. Our counterfactual experiments show that preferential attachment is a sufficient condition for the \(\beta\) coefficient exceeding phenomenon. This finding addresses~\citep{gomez2022leading}'s question, suggesting that higher paper counts from core countries boost visibility, thereby limiting the \(\beta\) coefficient’s effectiveness.
To validate the citational distortion theory, we conduct counterfactual experiments to demonstrate that, although core countries exhibit significantly higher SCR than peripheral countries, this disparity diminishes when the distribution of authors by country is held constant. This finding suggests that the uneven nationality distribution of researchers contributes to the elevated SCR observed in core countries. Furthermore, we confirm the absence of country-wise citational distortion and propose the RPS indicator to investigate its underlying mechanisms, validating that preferential referencing for papers from either core or peripheral countries is absent.

CiteAgent also serves as a social experiment platform with customizable academic environments with different experimental conditions, enabling the observation and analysis of human reference behaviors. Through repetitive simulation experiments within CiteAgent, we successfully replicate two dominant properties in network evolution: densification and shrinking diameter. 
We also successfully simulate the statistical results of traditional CCA analysis and extend it by using LLM-SE to examine citation importance levels, categorized by citation motivation and placement sections. 
Additionally, by designing academic experiments with single and multiple authors in CiteAgent, we illustrate how promoting academic communication can effectively enhance author creativity, providing valuable insights for real-world academic environments.

However, CiteAgent also has its limitations. One notable issue is that simulation authenticity is largely dependent on the human behavior simulation capability of LLMs. For instance, \texttt{GPT-3.5} struggles to simulate preferential referencing behavior motivated by citation-related information during reference selection. This indicates that untrained LLMs possess restricted simulation capabilities. 
The other limitation is that current modeling involves simplifications, particularly in considering only basic attributes such as nationality and research topic, while excluding factors like academic reputation and institutional influence~\citep{petersen2014reputation,bai2020measure}. 
Nonetheless, we believe that ongoing advancements in LLMs will enhance their capacity to simulate complex human behaviors. In the future, we aim to extend the LLM-SE and LLM-LE research paradigms to other domains of network science, and establish a reliable and scalable experimental platform for future simulation-based studies.

\section{Methods}

Our research presents a novel approach that integrates social consensus from LLMs to create a customizable experimentation platform, CiteAgent, for the science of science research. 
Built upon the AgentScope framework~\cite{agentscope}, CiteAgent enables scalable, multi-agent simulations of citation behaviors and knowledge diffusion. 
Below, we outline the execution workflow of the CiteAgent in detail.

Before simulation steps begin, CiteAgent undertakes three configuration setups:
\textbf{(1) Time Flow:} CiteAgent sets each simulation round to a fixed real-world duration, typically set to 5 days.
\textbf{(2) Seed Citation Network:} An initial citation network, \( G_0 \), is constructed using data from real-world citation networks. Given that graphs are mathematical structures of entity-wise interactions~\citep{bergmeister2024efficient}, \(G_0\) encompasses two entity types: author and paper. We collect seven text-rich attributes for papers and five for authors. The authors and papers in \(G_0\) form the initial author set \(A\) and paper collection set \( P \), respectively.
\textbf{(3) Paper Hyperparameter:} We set three parameters for paper creation in each simulation round: the number of new papers to be generated, denoted as \(n_p\); the number of recommended authors for each paper, denoted as \(n_a\); and the number of recommended citations for each paper, denoted as \(n_k\).

After setup configuration with \(G_0\), CiteAgent conducts iterative simulation steps, each simulation round typically representing a five-day period. As shown in Figure~\ref{fig1}, each simulation step integrates new papers into \(P\) and new authors into \( A\), following three main stages: \textbf{Initialization, Socialization, and Creation.} 

\textbf{(1) Initialization Stage}: This stage involves updating author set $A$ with heterogeneous LLM-based agents. To achieve a believable simulation of human reference behaviors, we adhere to a general paradigm~\citep{park2023generative} for constructing LLM-based agents. Having initialized set \( A \) with authors from \( G_0 \), we aim to further expand the quantity and diversity of authors in CiteAgent. 
We update $A$ by incorporating new authors and pruning overloaded ones, with the adjustments guided by the authors' activity frequency. 
Typically, we add 20 new authors every 5 days, and we designate an author as overloaded if his published papers reach the threshold (typically 100). The newly added authors are initialized with synthetic author profiles generated by LLMs, including nationality, expertise, institution, etc. Each author is designed to emulate human-like behaviors such as thinking, socializing, and drafting, and each author possesses a memory component to store their action trajectory. Detailed prompt templates used at each stage of this process are provided in Section 1 of the supplementary material.

\textbf{(2) Socialization Stage}: 
Moving on to the socialization stage, we select a subset of authors as the first authors for the created paper, denoted as \( A^a \). Suppose the number of new paper drafts is $n_p$, then the number of first authors is set to $|A^a| = n_p$. We aim for the first authors to engage in spontaneous academic communication with collaborators. To achieve this, first author \(a_i\) must first identify his collaborators. 
To assist in this process, we propose the Collaborator Recommendation Algorithm (CRA), which aims to recommend a set of collaborators \( A^{c}\)  from similar institutions, countries, or areas of academic expertise. 

Before recommending collaborators, we establish a threshold \(n_{a}\) for the number of recommended authors of each paper.
CRA then recommend \(A_i^{c}\) in two steps, as outlined in Algorithm \ref{alg:collaborator_recommendation}:
In the first step, the CRA recommends collaborators based on \(a_i\)'s collaboration history. For the authors that have collaborated with \(a_i\), we denote this collaborated author set as $A_i^{H}$.
We apply Breadth-First Search (BFS) to select collaborators from \(A_i^{H}\) until the queue is empty or the number of collaborators \(| A_i^{c} | \geq n_{a}\).

If \(|A_i^{c}| < n_{a}\), we proceed to the second step. In this step, we filter \(A_i^{H}\) based on the author profile of \(a_i\). The system pre-defines a list of author attributes for CRA, denoted as \(PF^{c}\). In our experiment, \(PF^{c}\) is arranged in the following order: Topic \(\rightarrow\) Expertise \(\rightarrow\) Citation \(\rightarrow\) Institution \(\rightarrow\) Nationality. For each attribute \(pf \in PF^{c}\), we filter \(a_j\) from \(A\) if \(a_j\) shares the same value as \(a_i\) for the attribute \(pf\). Through multiple iterations of this process, we gradually add collaborators to \(A_{i}^{c}\).

\begin{algorithm}
  \caption{Collaborator Recommendation Algorithm for Author $a_i$}
  \label{alg:collaborator_recommendation}
  \begin{algorithmic}[1]
      \State \textbf{Input:} Author $a_i$, Collaboration History $A_{i}^{H}$, Threshold $n_{a}$, Author Profile Filters $PF^{c}$
      \State \textbf{Output:} Recommended Collaborators $A_{i}^{c}$
      \State $A_{i}^{c} = \emptyset$
      \State \textbf{Step 1: Filter Based on Collaboration History}
      \State Initialize queue $Q \gets A_{i}^{H}$
      \While{$| A_{i}^{c} | < n_{a}$ \textbf{ and } $Q \ne \emptyset$}
          \State $a_j \gets Q$.pop()
          \State $A_{i}^{c}$.append($a_j$)
          \For{$a_k \in A_j^H$}
              \If{$a_k \notin A_{i}^{c}$ \textbf{ and } $a_k \notin Q$}
                  \State $Q$.append($a_k$)
              \EndIf
          \EndFor
      \EndWhile
      \\
      \State \textbf{Step 2: Filter Based on Author Profile}
      \While{$| A_{i}^{c} | < n_{a}$}
          \For{$pf$ in $PF^{c}$}
            \For{$a_j$ in $\left\{a \in A \mid p f(a)=p f\left(a_i\right)\right\}$}
              \State $A_{i}^{c}$.append($a_j$)
            \EndFor
          \EndFor
      \EndWhile
    % \Return $A_{i}^{c}$
  \end{algorithmic}
\end{algorithm}

After the collaborator recommendation, the first author \(a_i\) selects desired collaborators from \(A_i^{c}\). Then, the first author $a_i$ initiates a discussion with the selected collaborators.
In our experiment setup, we set the authors to discuss for two rounds for the CiteSeer and Cora datasets, and one round for the LLM-Agent dataset. After the discussion, the first author generates the draft of the paper. 

\textbf{(3) Creation Stage}: 
To ensure that the author can produce high-quality papers, creation stage is divided into three steps: preference update, reference paper search, and paper writing. 

First, we update the author profile of $a_i$ in each simulation round, which is conducted by the memory component $m_{i}$.
To ensure authors can produce high-quality papers within CiteAgent, we equip \(a_i\) with a memory component \(m_{i}\) that captures the author's action history. We implement two types of memory: social memory and writing memory. As mentioned above, the former is used to store the discussion histories with collaborators, while the latter is used to store the author's paper-writing history. We utilize reflection and summarization techniques~\citep{shinn2023reflexion} to organize the memory stream. At the beginning of each paper creation stage, we update the memory stream \(m_{i}\). 
% Then, the first author $a_i$ initiates a discussion with the selected collaborators.
% After the socialization stage, we update the memory stream of $a_i$. 
% In our experiment setup, we set the authors to discuss for two rounds for the CiteSeer and Cora datasets, and one round for the LLM-Agent dataset. After the discussion, the first author generates the draft of the paper. After the socialization stage, we update the memory stream of $a_i$. 

Next, To simulate the referencing paper search process in real-world academic environments, we develop a scholarly search engine.
When first author \( a_i \) conducts a search to retrieve the most relevant papers, he first generate a query set \( Q_i \), which is conducted with a list of query words about paper keywords and topics. To get recommended papers \( P_i^c \) based on query words, the process is divided into two steps as outlined in Algorithm \ref{alg:scholar}: recall and rerank.
Firstly, For each query word \( q \in Q_i \), in the recall stage, we filter the top papers using a vector retriever. We begin by constructing a vector database from the embedding vectors of existing papers \( P \). Using the \texttt{SBERT}~\citep{reimers-2019-sentence-bert} as our embedding model, we embed each paper \( p \in P \) into a 384-dimensional vector \( e_p \). 
we embed \( q \) as a vector \( e_q \) and calculate the cosine similarity between \( e_q \) and \( e_p \) to identify the most relevant papers. This process recommends the top 20 most relevant paper set for each \( q \), denoted as \( o_i^q \).
Then, in the rerank stage, we reorder \( o_{i}^q \)based on the personal preference of authors, producing a final candidate paper set \( P_{i}^{c} \). Similar to CRA, the system predefines a list of paper attributes for the paper recommendation, denoted as \(PF^{p}\). 
For each attribute \(pf \in PF^{p}\), we rearrange the paper order in \(o_i^q\) based on the attribute \(pf\). In our experiment, \(PF^{p}\) is arranged in the following order: Paper Citation $\rightarrow$ Paper Topic. As a result, \( o_{i}^q \) is first sorted based on citation counts, and then by paper topics.

Finally, we aggregate all \( o_i^q \) entries into a combined set. Due to the context length constraints of LLMs, we typically select the top 20 papers from the combined set to form the final candidate paper set, \( P_i^c \). 
The results of \( P_{i}^{c} \) are returned to \( a_i \). Ultimately, \( a_i \) refines paper draft by referencing papers from \( P_{i}^{c} \), adhering to the citation limit \( n_k \). As a result, \( n_p \) first authors collectively write \( n_p \) papers in one simulation round.

\begin{algorithm}
\caption{Paper Search by Scholarly Search Engine for Author $a_i$.}
\label{alg:scholar}
\begin{algorithmic}[1]
  \State \textbf{Input:} Author $a_i$, Query set $Q_i$, Paper Profile filters $PF^{p}$
  \State \textbf{Output:} Candidate Paper Set $P_{i}^{c}$
\State $P_{i}^{c} = \emptyset$,
\For{$q \in Q_i$}
    \State $\operatorname{top20}_{p \in P} \operatorname{Sim}\left(q, p\right) \rightarrow o_i^q \quad \text{where} \quad
    \operatorname{Sim}\left(q, p\right) = \frac{e_q \cdot e_{p}}{\left\|e_q\right\| \cdot\left\|e_{p}\right\|}$,
    \For{$pf \in PF^{p}$}
      \State $o_i^q= $ \Call{ReRank}{$o_i^q, pf$},
    \EndFor
    \State $P_{i}^{c}$.append$(o_i^q)$,
\EndFor
\State $P_{i}^{c}$ = $\operatorname{top20}(P_{i}^{c})$
\end{algorithmic}
\end{algorithm}

By iterating through the above three stages, we can progressively enrich the paper database \( P \), realistically modeling the evolutionary process of the citation network.
Given authentic simulation data from CiteAgent, we can perform in-depth analyses for science-of-science research using LLM-based research paradigms. 

Within the LLM-LE research paradigm, we first conduct a series of counterfactual experiments in customized academic environments. We control the information visibility of \(P_c\), including paper content, citation, timeliness, author.
By gathering the results of these counterfactual experiments, we examine the citation network structure, including power-law distributions, citational distortion, decreasing diameter and etc.

On the other hand, we adopt the LLM-SE research paradigm to explore the psychological motivations behind human behavior. Traditional studies have largely relied on methods such as questionnaires, surveys~\citep{case2000can}, and small-scale social experiments~\citep{fowler2014motivating}, focusing on the interplay between biological predispositions in physical environments with constraints and affordances. However, these approaches often struggle to capture real-time psychological data as individuals engage in different actions. In contrast, LLMs, guided by RLHF training~\citep{christiano2017deep}, can be prompted to share their authentic thoughts during actions. Thus, we prompt authors to identify their key psychological motivations regarding reference selection, citation placement, and perceived importance, facilitating a more nuanced analysis of the psychological motivations that drive human behavior patterns.

\section{Data Availability}

All data that were used to create the figures are available on the Github Repository at
\url{https://github.com/Ji-Cather/CiteAgent}.

\section{Code Availability}
All code that were used in experiments are available on the Github Repository at
\url{https://github.com/Ji-Cather/CiteAgent}.

\section{Funding Statement}
This research was supported in part by National Natural Science Foundation of China (No. U2241212, No. 92470128), by National Science and Technology Major Project (2022ZD0114802), by Beijing Outstanding Young Scientist Program No.BJJWZYJH012019100020098. We also wish to acknowledge the support provided by the fund for building world-class universities (disciplines) of Renmin University of China, by Engineering Research Center of Next-Generation Intelligent Search and Recommendation, Ministry of Education, by Intelligent Social Governance Interdisciplinary Platform, Major Innovation \& Planning Interdisciplinary Platform for the “Double-First Class” Initiative, Public Policy and Decision-making Research Lab, and Public Computing Cloud, Renmin University of China.

\section{Ethics Declarations}

\noindent\textbf{Competing interests} 

\noindent The authors declare no competing interests.

\noindent\textbf{Ethical approval}

\noindent This article does not contain any studies with human participants performed by any of the authors.

\noindent\textbf{Informed consent}

\noindent This article does not contain any studies with human participants performed by any of the authors.

\bibliography{sn-bibliography}% common bib file

\end{document}